\documentclass[a4paper]{PoS}

\usepackage{textcomp}
\usepackage{amsfonts} 
\usepackage{amssymb} 
\usepackage{amsmath} 
\usepackage{slashed}
\usepackage{graphicx}
\usepackage{pifont}
\usepackage{multirow,lscape}
\usepackage{array}
\usepackage{longtable}
\usepackage{verbatim}
\usepackage{bm}
\usepackage{subfigure}
\usepackage[usenames,dvipsnames]{xcolor}
\usepackage{url} 
\usepackage{comment}

\graphicspath{{./figs/}}

\allowdisplaybreaks

\title{
Heavy-heavy current improvement for calculation of $\bar{B}\rightarrow
D^{(*)}\ell \bar{\nu}$ semi-leptonic form factors using the 
Oktay-Kronfeld action
}

\ShortTitle{Heavy-heavy current improvement}

\author{ \speaker{Jon~A.~Bailey\footnote{Speaker.}, Jaehoon~Leem}, Weonjong~Lee\\
  Lattice Gauge Theory Research Center, CTP, and FPRD, \\
  Department of Physics and Astronomy, \\
  Seoul National University,
  Seoul 08826, South Korea\\
  E-mail: \email{jonbailey@snu.ac.kr}, 
  \email{leemjaehoon@gmail.com},
  \email{wlee@snu.ac.kr} }

\author{Yong-Chull~Jang \\
        Los Alamos National Laboratory, \\
        Theoretical Division T-2, MS B283, \\
        Los Alamos, New Mexico 87545, USA \\
        E-mail: \email{integration.field@gmail.com} }
\author{SWME Collaboration}

\abstract{
Lattice calculations of the form factors for $\bar{B}\to D^{(*)}\ell\bar{\nu}$
decays can be used to extract the CKM matrix element $|V_{cb}|$.  
The Oktay-Kronfeld action is a highly improved version of the Fermilab action,
which systematically reduces heavy quark discretization effects through
$\mathcal{O}(\lambda^3)$ in HQET power counting, for heavy-light meson
quantities.
To calculate $\bar{B}\rightarrow D^{(*)}\ell \bar{\nu}$ semi-leptonic 
form factors using Oktay-Kronfeld heavy quarks, we need to improve the 
heavy quark currents to the same level.
We report our progress in calculating the improvement coefficients
for currents composed of bottom and charm quarks.
Our results presented in this paper are preliminary.
}

\FullConference{34th Annual International Symposium on Lattice Field Theory\\
		24-30 July 2016\\
		University of Southampton, UK}

\renewcommand{\vec}{\bm}

\begin{document}

\section{\label{sec:intro}Introduction}

Precision tests of the Standard Model (SM) in the flavor sector are crucial to
the search for new physics.
The SM requires unitarity of the Cabibbo-Kobayashi-Maskawa (CKM) quark mixing
matrix, and the CKM matrix element $|V_{cb}|$ plays a central role because it
normalizes the unitarity triangle.
By calculating $\bar{B}\rightarrow D^{(*)}\ell\bar{\nu}$ 
decay form factors on the lattice and combining them with 
experimental results for the branching fractions, one can 
determine $|V_{cb}|$.
The decay rates for these processes are given by
\begin{align}
\frac{d\Gamma}{dw}(\bar{B}\rightarrow D \ell \bar{\nu})
& = \frac{G^2_F |V_{cb}|^2m_B^5}{48\pi^3}
(w^2 -1 )^{3/2}r^3 (1+r)^2 
|\eta_{\text{EW}}|^2 |\mathcal{F}_D(w)|^2,
\\
\frac{d\Gamma}{dw}(\bar{B}\rightarrow D^* \ell \bar{\nu})
& = \frac{G^2_F |V_{cb}|^2m_B^5}{48\pi^3} 
(w^2 -1 )^{1/2}r^{*3} (1-r^*)^2 
\nonumber \\ 
& \times \left[ 1+ \frac{4w}{w+1} \frac{1-2wr^* + r^{*2}}{(1-r^*)^2} \right] 
|\eta_{\text{EW}}|^2
|\mathcal{F}_{D^*}(w)|^2,
\end{align}
where $w = v_B \cdot v_{D^{(*)}}$ is the recoil parameter, $r^{(*)} =
m_{D^{(*)}}/m_B$ are the ratios of the daughter to parent meson masses,
$\eta_{\text{EW}}$ incorporates higher order electroweak corrections, and
$\mathcal{F}_{D^{(*)}}(w)$ are the hadronic form factors.

The most precise results for
$|V_{cb}||\eta_{\text{EW}}||\mathcal{F}_{D^{(*)}}(1)|$ from experimental
measurements by BABAR \cite{Aubert:2008yv,Aubert:2007qs, Aubert:2007rs} and
Belle \cite{Dungel:2010uk} have uncertainties of a few percent.
Recently the Fermilab-MILC Collaboration extracted $|V_{cb}|$ \textit{via} lattice
calculations of the form factors for $\bar{B}\rightarrow D^{*}\ell \bar{\nu}$
at zero recoil \cite{Bailey:2014tva} and $\bar{B}\rightarrow D\ell \bar{\nu}$
at nonzero recoil~\cite{Lattice:2015rga,DeTar:2015orc}.
The results have $2-5$\% total uncertainties and are consistent with one
another and the determination of the $\bar{B}\to D\ell\bar{\nu}$ form factors by
the HPQCD Collaboration~\cite{Na:2015kha}. 
If we can reduce the discretization errors in lattice calculations of the form
factors, then we can determine $|V_{cb}|$ to higher precision.

The Fermilab-MILC Collaboration used the Fermilab action for the $b$ and $c$
quarks.
The Fermilab improvement program controls lattice 
cutoff effects at any quark mass \cite{ElKhadra:1996mp}.
To reduce heavy-quark discretization errors, Oktay and Kronfeld extended the
improvement of the Fermilab (clover) action to higher order, including mass
dimension 6 and 7 operators, or through third order in HQET power counting.
Tests of the tree-level matched Oktay-Kronfeld (OK) action yield promising
results~\cite{MBO:LAT2010,Bailey:2014zma,Bailey:2016kbw}.
For systematic improvement of form factor calculations, the flavor-changing
currents must be improved to the same level.
The authors of Ref.~\cite{ElKhadra:1996mp} defined improved currents in terms
of an improved heavy quark field.  
We begin our construction of improved currents by extending the improved field
to include operators corresponding to mass dimensions 5 and 6.

In Ref.~\cite{Bailey:2014jga} we considered the improvement of two-quark matrix
elements of the flavor-changing currents.  
We showed that these matrix elements can be matched through third order in
expansions of the heavy quark momenta and wrote down results for four of the
11 improvement parameters entering the improved heavy quark fields.
In Sec.~\ref{sec:improved_current} we further discuss current improvement and
the improved heavy quark field.
We present details of the matching process and results for the remaining seven
improvement parameters in Sec.~\ref{sec:matching_cal}.  Section~\ref{sec:end}
contains a status summary and outstanding issues.

\section{\label{sec:improved_current}Current improvement}

The mass-dependent renormalization program begins with the observation that
improved Wilson actions can be tuned to the renormalized trajectory by lifting
the constraint of time-space axis interchange symmetry, including only
irrelevant operators that do not lead to modifications of the Wilson time
derivative.  The resulting class of actions are constructed to approach the
continuum limit for any quark mass, including quark masses large in lattice
units.  

Consequently, the improvement parameters are functions of the quark masses,
\textit{i.e.}, mass-dependent~\cite{ElKhadra:1996mp}.  In this context one may use suitable
generalizations of the Symanzik action, HQET, or NRQCD as effective continuum
field theories to describe the physics of the resulting lattice theory,
including cutoff
effects~\cite{Kronfeld:2002pi,Oktay2008:PhysRevD.78.014504,Kronfeld:PhysRevD.62.014505}.
These descriptions are useful for designing improved actions and currents,
assessing improvement, and quantifying remaining (systematic) discretization
effects~
\cite{Oktay2008:PhysRevD.78.014504,Kronfeld:PhysRevD.62.014505,Harada:2001fi,Harada:2001fj}.

The Oktay-Kronfeld action was designed to reduce charm quark discretization
errors to less than about $1\%$; bottom quark discretization errors are even
smaller~\cite{Oktay2008:PhysRevD.78.014504}.  Improvement of the heavy quark
currents proceeds in the same way, in principle:  All operators with the
quantum numbers of the desired currents are included, through a given order in
the power counting~\cite{ElKhadra:1996mp},
\begin{align}
J_\mathrm{QCD} = Z_J(am_q,g^2) \left[ J_0 + \sum_i C_i(am_q,g^2) J_i \right]\,.
\end{align}
The coefficients $C_i$ of the improvement operators $J_i$, together with the
renormalization factors $Z_J$, are fixed by matching matrix elements to their
continuum values~\cite{ElKhadra:1996mp}.  The matching can be carried out using
the power counting of the effective continuum field theory to systematize
improvement, and for arbitrary fermion
masses~\cite{ElKhadra:1996mp,Harada:2001fi,Harada:2001fj}.

In practice, for improvement through first order in HQET, introducing an
improved heavy quark field suffices.  The improved current can be written 
\begin{align}
J_\Gamma = \bar{\Psi}_{Ic} \Gamma \Psi_{Ib}\,,\label{eq:impr_current}
\end{align}
where $\Gamma$ indicates the Dirac structure, and $\Psi_{If}$ is the improved
heavy quark field for flavor $f = c,b$,
\begin{align}
\Psi_{If}(x) = e^{M_{1f}/2}(1 + d_{1f} \bm{\gamma \cdot D})\psi_f(x)\,,\label{eq:impr_field}
\end{align}
where $M_{1f}$ is the tree-level rest mass of $f$ quarks, $d_{1f}$ is the
improvement parameter, which depends on this rest mass, $\bm{D}$ is the lattice
(symmetric) covariant derivative, and $\psi_f$ is the heavy quark field
appearing in the mass form of the action.  (Unless explicitly indicated, we set
the lattice spacing $a = 1$.)
The results for the improvement parameters $d_{1f}$ at tree-level were first
written down in Ref.~\cite{ElKhadra:1996mp}; these results were obtained by
tree-level matching of two-quark matrix elements of the flavor-changing
currents.  
The authors of Refs.~\cite{Harada:2001fi,Harada:2001fj} then showed
that the current defined by Eqs.~\eqref{eq:impr_current} and
\eqref{eq:impr_field} is improved through first order in HQET power
counting.

For improvement of two-quark matrix elements through third order in the momenta
of the heavy quarks, an improved field again suffices, with the ansatz
~\cite{Bailey:2014jga}
\begin{align}
\Psi_I(x) =& e^{M_1/2}\bigg[1+d_1 \vec{\gamma} \cdot \vec{D}
          +\frac{1}{2}d_2 \Delta^{(3)}
          + \frac{1}{2}i d_B  \vec{\Sigma} \cdot \vec{B}
          + \frac{1}{2}d_E  \vec{\alpha} \cdot \vec{E} \nonumber \\
         &+ d_{r_E} \{ \vec{\gamma} \cdot \vec{D}, \vec{\alpha} \cdot \vec{E}\}
          + d_{z_E} \gamma_4(\vec{D} \cdot \vec{E}
          - \vec{E} \cdot \vec{D}) \nonumber \\
         &+ \frac{1}{6}d_{3} \gamma_i D_{i} \Delta_{i}
         + \frac{1}{2}d_{4}\{\vec{ \gamma} \cdot \vec{D},\Delta^{(3)} \}
          + d_{5}\{\vec{ \gamma} \cdot \vec{D} ,i \vec{ \Sigma} \cdot \vec{B}\}\nonumber \\
         &+ d_{EE}\{\gamma_4 D_{4 }, \vec{ \alpha} \cdot \vec{E}\}
          + d_{z_3} \vec{\gamma} \cdot (\vec{D}\times \vec{B}
          + \vec{B}\times \vec{D}) \bigg] \psi(x)\,.
\label{eq:field-imp}
\end{align}
Matching two-quark matrix elements at tree-level yields results for the
parameters $d_1$, $d_2$, $d_3$, and $d_4$~\cite{Bailey:2014jga}.  
The remaining seven improvement parameters in this ansatz do not enter
tree-level calculations of the two-quark matrix elements and so are not
determined by tree-level matching calculations of these matrix elements.
To determine these parameters we match four-quark matrix elements of the
currents, at tree-level, as described below.

\section{\label{sec:matching_cal}Matching}

To obtain improvement parameters through third order in HQET, we consider the
following four-quark matrix element of flavor-changing currents,
\begin{align}
\langle \ell(\eta_2, p_2) u(\eta{'}, p{'})|
\bar{\psi}_{u}
& \Gamma \Psi_{b} | b(\eta, p) \ell(\eta_1, p_1)\rangle_{\text{lat}}\,,\label{eq:4qme}
\end{align}
where $\ell$ represents a light spectator quark, and $u$ and $b$ indicate an up
quark and a bottom quark, respectively.  We consider the transition to the
light (up) quark rather than to the heavy (charm) quark to simplify the
calculation; diagrams with improvement terms on the daughter quark line are
eliminated.  Accordingly, the field $\psi_u$ appears in Eq.~\eqref{eq:4qme}.

\begin{figure}[t!]
\centering
\subfigure[]{
\label{fig:cont-diagram_1}
\includegraphics[width=0.45\textwidth]{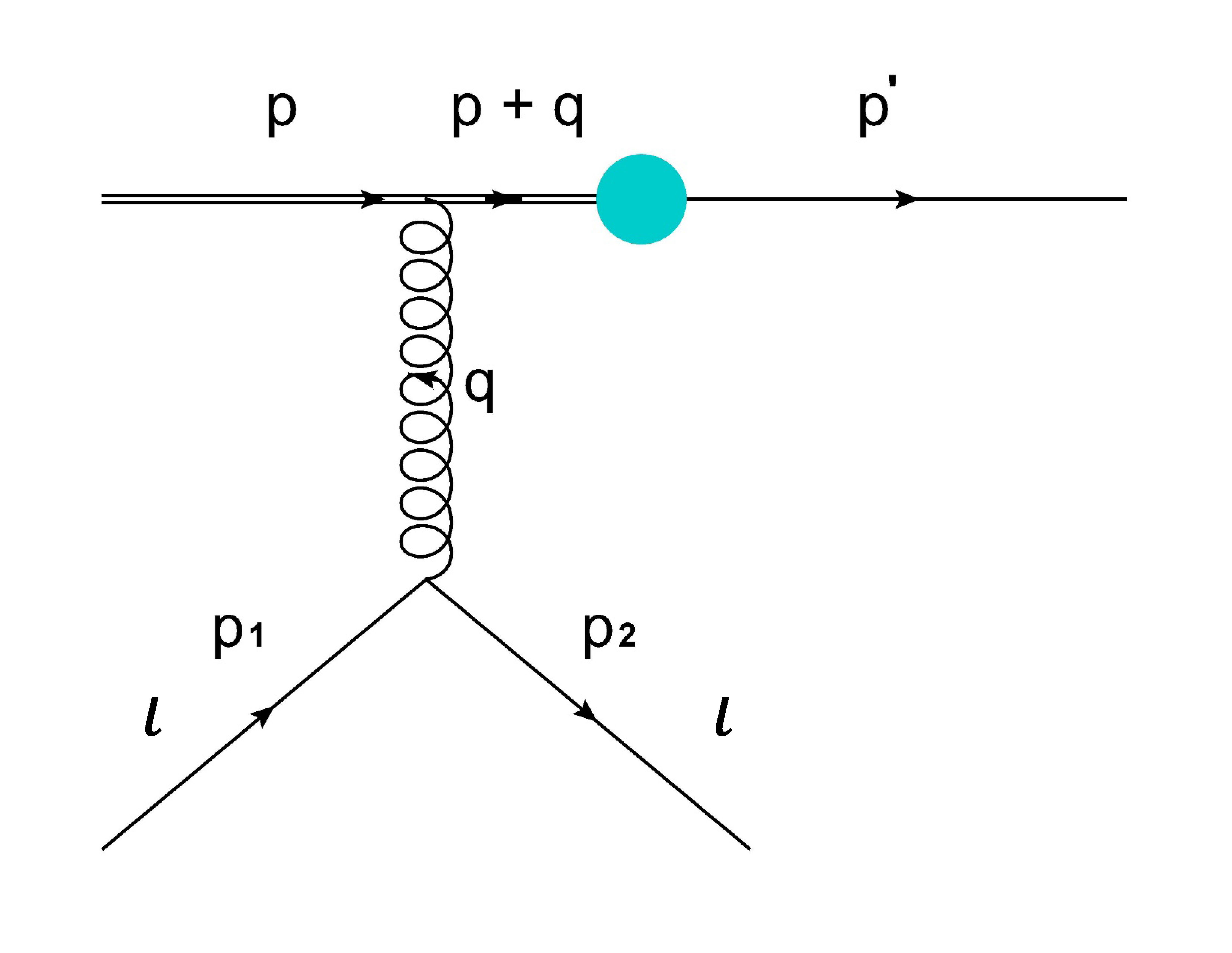}
}
\subfigure[]{
\label{fig:cont-diagram_2}
\includegraphics[width=0.45\textwidth]{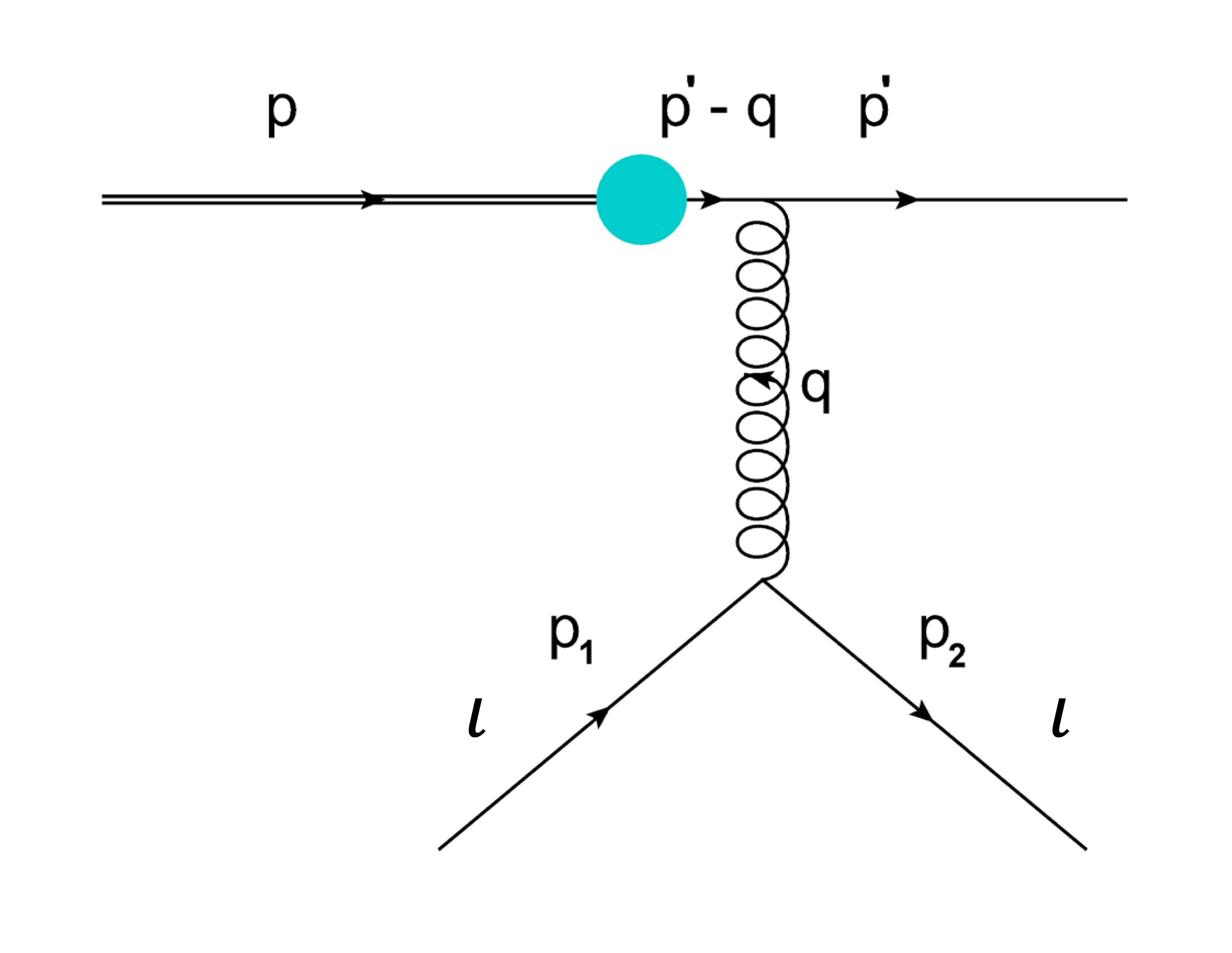}
}
\caption{Tree-level, continuum diagrams for 
the four-quark matrix element.
The colored circle represents the flavor-changing current. 
(a) The diagram with gluon exchange at the heavy-quark line.
(b) The diagram with gluon exchange at the light-quark 
line.}
\end{figure}

The corresponding continuum, tree-level diagrams are given in
Figs.~\ref{fig:cont-diagram_1} and~\ref{fig:cont-diagram_2}.
The gluon exchange may occur at the external line of the $b$-quark or the
$u$-quark.
In the case of gluon exchange at the $u$-quark line, however, the matching
condition from the diagram is equivalent to that from the two-quark matrix
element mentioned above.

\begin{figure}[t!]
\centering
\subfigure[]{
\label{fig:latt-diagram_1}
\includegraphics[width=0.45\textwidth]{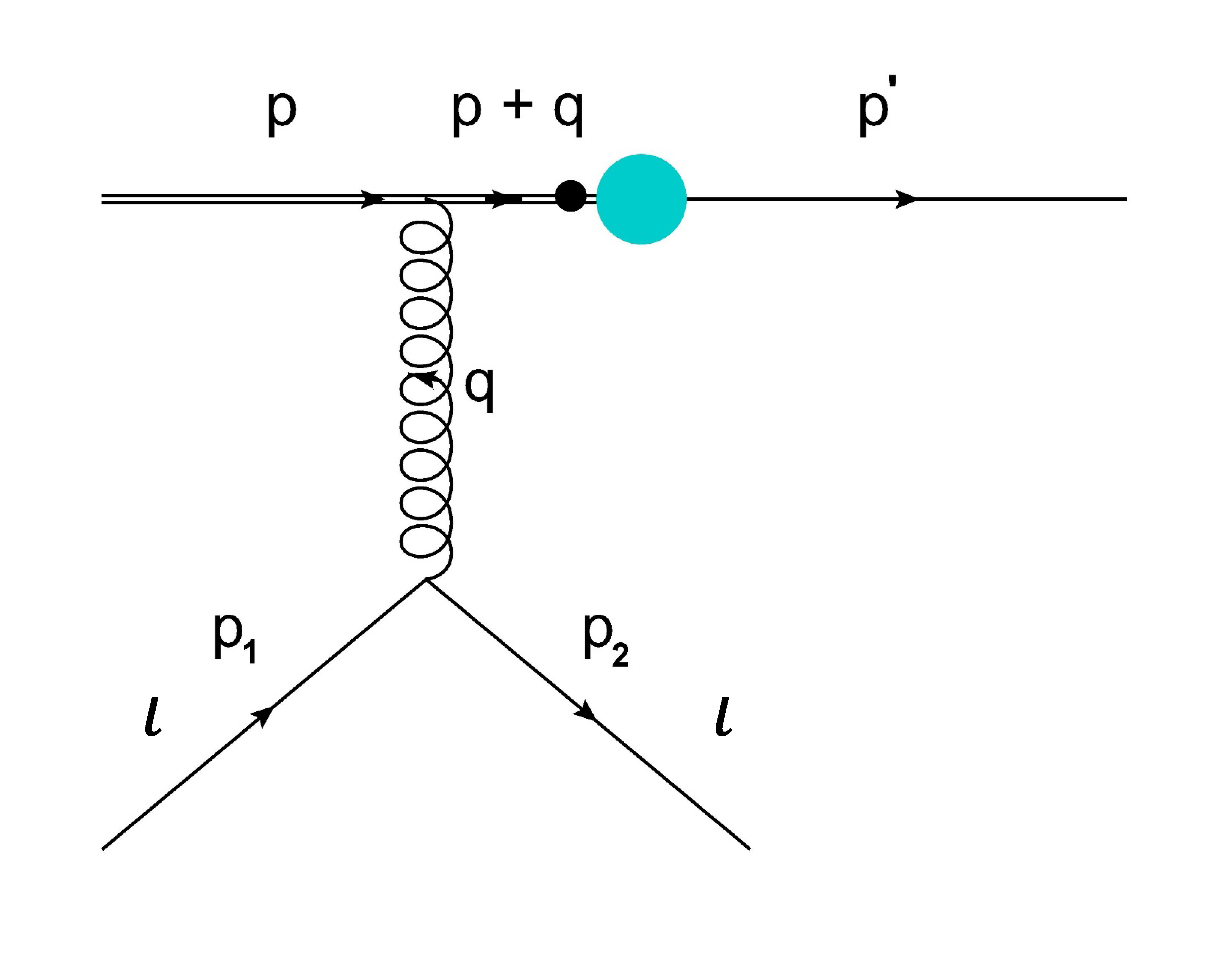}
}
\subfigure[]{
\label{fig:latt-diagram_2}
\includegraphics[width=0.45\textwidth]{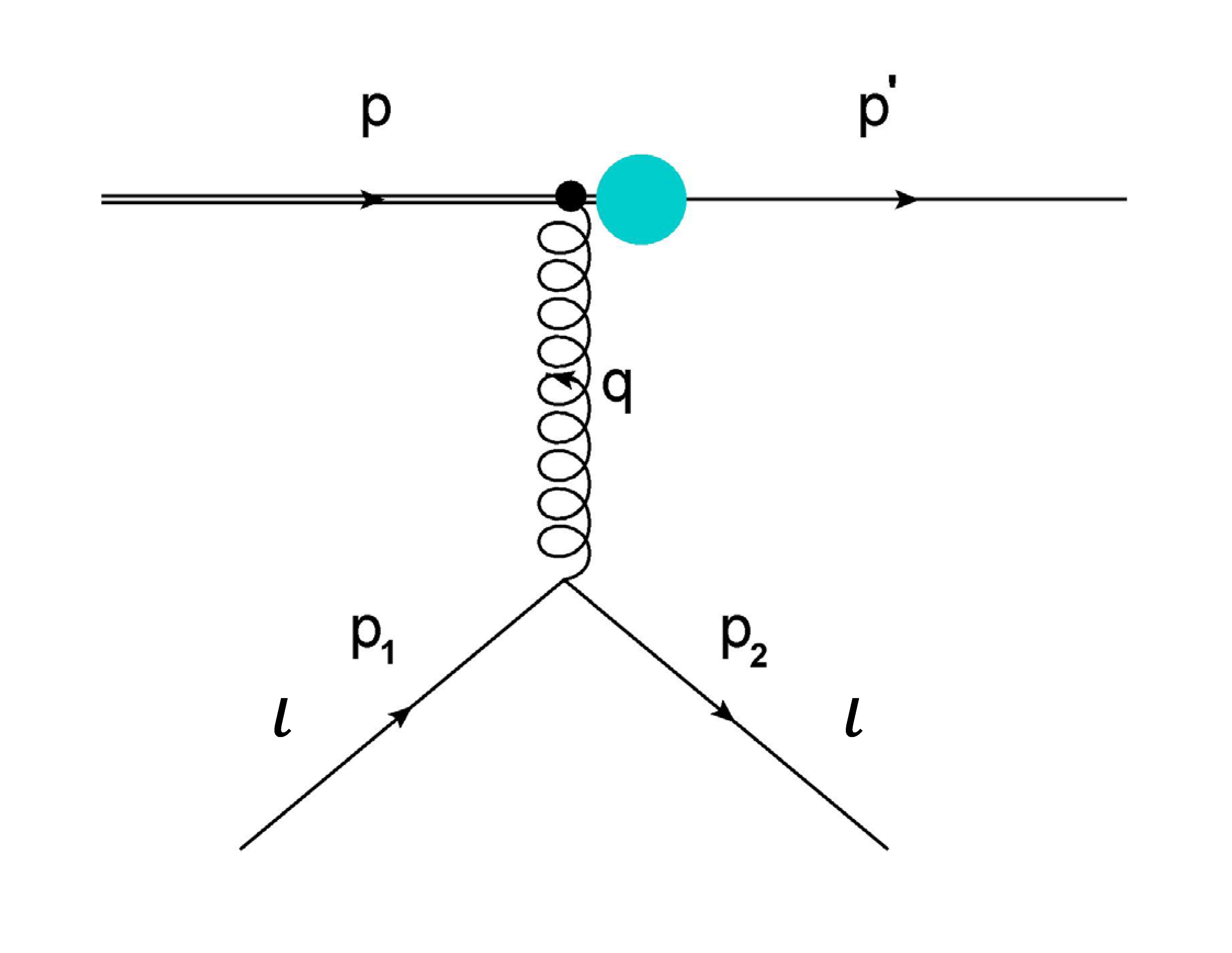}
}
\caption{Tree-level, lattice diagrams with gluon exchange 
at the heavy-quark line. 
The dots represent (a) the zero-gluon vertex and (b) the
one-gluon vertex from the improved quark field, 
respectively.}
\end{figure}
Therefore, the part we need to consider is the one-gluon exchange vertex at the
$b$-quark line.
The corresponding lattice diagrams are given in Figs.~\ref{fig:latt-diagram_1}
and~\ref{fig:latt-diagram_2}.
In addition to one-gluon exchange from the action vertices
[Fig.~\ref{fig:latt-diagram_1}], one-gluon exchange occurs from vertices of the
improvement terms in the quark field [Fig.~\ref{fig:latt-diagram_2}]; these
vertices are represented by the dots in Figs.~\ref{fig:latt-diagram_1}
and~\ref{fig:latt-diagram_2}.
The corresponding sub-diagrams in the continuum and lattice
theories are
\begin{align}
\mathcal{M}_\mu
\label{eq:matching-vertex1}
&=(-gt^a)S_b(p+q)
\gamma_\mu \sqrt{\frac{m_b}{E_b}}u_b(\eta,p)
=(-gt^a) \frac{m_b-i\gamma \cdot(p+q)}{m_b^2+(p+q)^2} \gamma_\mu 
\sqrt{\frac{m_b}{E_b}}u_b(\eta,p),
\\
\label{eq:matching-vertex2}
\mathcal{M}^{\text{lat}}_\mu
&=(-gt^a)n_\mu(q)
\Big[ R^{(0)}_b(p+q)  S^{\text{lat}}_b(p+q)\Lambda_\mu(p+q,p)
+R^{(1)}_{b,\mu}(p+q,p)\Big]
\mathcal{N}_b(p)u^{\text{lat}}_b(\eta,p),
\end{align}
where $q$ is the momentum of the virtual gluon; $p$ is the external
$b$-quark momentum; the factor $n_\mu(q) = 2\sin
\left(\dfrac{1}{2}q_\mu \right) / q_\mu = 1- \dfrac{1}{24}q_\mu^2
+\cdots$ is the lattice wave-function factor for the gluon line;
$S_b(p+q)$ and $S^{\text{lat}}_b(p+q)$ are the continuum and lattice
$b$-quark propagators, respectively;
$\mathcal{N}_b(p)u^{\text{lat}}_b(\eta,p)$ is the normalized lattice spinor
from the OK action, which corresponds to the normalized spinor
$\sqrt{\dfrac{m_b}{E_b}}u_b(\eta,p)$ in the continuum;
$\Lambda_\mu$ denotes the one-gluon exchange vertex from the OK
action; $R^{(0)}_b$ denotes zero-gluon vertices from the improvement
terms in the heavy quark field; and $R^{(1)}_{b,\mu}$ denotes
one-gluon vertices from the improvement terms in the heavy quark
field.

The improvement parameters $d_1$, $d_2$, $d_3$, and $d_4$ enter
\textit{via} the zero-gluon exchange vertices $R^{(0)}_b$,
\begin{align}
&R^{(0)}_b(p+q) 
=e^{M_1/2} \Bigg[
1+id_1 \sum_j \gamma_j \sin(p_j +q_j) - 2d_2\sum_j \sin^2 \frac{1}{2}(p_j+q_j)
\nonumber \\
&-\frac{2}{3}i d_3 \sum_j \gamma_j \sin(p_j+q_j)\sin^2 \frac{1}{2}(p_j+q_j)
-4i d_4 \sum_{j,k}\gamma_j \sin(p_j+q_j) \sin \frac{1}{2}(p_k+q_k)
\Bigg]\,.
\end{align}
The remaining seven improvement parameters enter \textit{via} the
one-gluon exchange vertices $R^{(1)}_{b,\mu}$.
The temporal ($\mu=4$) component is
\begin{align}
&R^{(1)}_{b,4} (p+q,p)
=e^{M_1/2}\cos \frac{1}{2}q_4 \gamma_4 \Bigg[
\frac{i}{2}d_E \zeta \sum_j \gamma_j \sin q_j
-d_{EE}\gamma_4 \sum_j \gamma_j \sin q_j
\big[\sin(p_4 +q_4) - \sin(p_4)\big]
\nonumber \\
&+(d_{r_E}-d_{z_E})\sum_j \sin q_j \big[\sin(p_j+q_j)-\sin p_j \big]
-id_{r_E}\sum_{j,l,m} \epsilon_{jlm} \Sigma_j \sin q_l \big[\sin(p_m+q_m) +\sin p_m\big]
\Bigg]\,,
\end{align}
while the spatial ($\mu=i$) components are
\begin{align}
&R^{(1)}_{b,i}(p+q,p)
=e^{M_1/2} \Bigg[-d_1  \gamma_i \cos (p_i +\frac{1}{2}q_i)
-id_2 \sin (p_i +\frac{1}{2}q_i)
-\frac{1}{2}d_B \sum_{r,m} \epsilon_{irm} \Sigma_m \sin q_r \cos \frac{1}{2}q_i
\nonumber \\
&-\frac{i}{2}d_E \gamma_4 \gamma_i \cos \frac{1}{2}q_i \sin q_4
+d_{r_E} \sum_{r,m}i\epsilon_{irm}\Sigma_m \gamma_4 \sin q_4 \big[\sin (p_r+q_r)+\sin p_r \big] 
\cos \frac{1}{2}q_i
\nonumber \\
&-(d_{r_E}-d_{z_E})\gamma_4 \sin q_4 \big[\sin (p_i+q_i)-\sin p_i \big] \cos \frac{1}{2}q_i
+d_{EE}\gamma_i \sin q_4  
\big[ \sin(p_4 +q_4) -\sin p_4 \big]\cos \frac{1}{2}q_i
\nonumber \\
&+\frac{1}{2}d_4 
\bigg[\gamma_i \cos(p_i +\frac{1}{2}q_i)\sum_j 4\big[ \sin^2 \frac{1}{2}(p_j+q_j)
+\sin^2 \frac{1}{2}p_j \big]
+2\sin(p_i +\frac{1}{2}q_i)\sum_j \gamma_j \big[\sin(p_j +q_j) +\sin p_j \big]
\bigg]
\nonumber \\
&+\frac{1}{12}d_3 \gamma_i \bigg[4\cos(p_i +\frac{1}{2}q_i)
\big[ \sin^2 \frac{1}{2}(p_i+q_i) +\sin^2 \frac{1}{2}p_i \big]
+2\sin(p_i +\frac{1}{2}q_i) \big[\sin(p_i +q_i) +\sin p_i \big]
\bigg]
\nonumber \\
&+(d_5-d_{z_3})\cos \frac{1}{2}q_i 
\Big[-\sum_j \gamma_j \sin q_j \big[ \sin(p_i +q_i) - \sin p_i \big]
+\gamma_i \sum_j \sin q_j \big[\sin(p_j+q_j)-\sin q_j\big]\Big]
\nonumber \\
&+d_5 \cos \frac{1}{2}q_i \gamma_4 \gamma_5 \sum_{r,m}\epsilon_{irm} 
\sin q_r \big[\sin(p_m+q_m) +\sin p_m \big]
\Bigg]\,.
\end{align}
In these expressions the momentum $p$ of the heavy quark is on shell,
$p_4 =iE_b = i\bigg[M_1+\dfrac{\boldsymbol{p}^2}{2M_2}+\cdots \bigg]$.

Assuming the spatial momentum $\boldsymbol{p}$ of the external $b$-quark and
the gluon momentum transfer $q$ are much smaller than the $b$-quark mass $m_b$
and the inverse of the lattice spacing $1/a$, we expand
$\mathcal{M}^{\text{lat}}_\mu$ and $\mathcal{M}_\mu$ in $\boldsymbol{p}a$,
$qa$, $\boldsymbol{p}/m_b$, and $q/m_b$.
We then equate the coefficients of the expansions of
$\mathcal{M}^{\text{lat}}_\mu$ and $\mathcal{M}_\mu$ at each order in
$\boldsymbol{p}$ and $q$.
The resulting equations constrain the improvement parameters $d_i$ from the
improvement vertices $R^{(0)}_b$ and $R^{(1)}_{b,\mu}$.
Each improvement parameter $d_i$ may appear in more than one
constraint equation.
The parameters $d_E$, $d_{r_E}$, $d_{z_E}$, and $d_{EE}$ appear in both the
temporal ($\mu=4$) and spatial ($\mu=i$) components of the one-gluon exchange
vertex $R^{(1)}_{b,\mu}$.
The constraint equations for $\mu=4$ and $\mu =i$ must be consistent.

For example, consider the constraint equation for the improvement parameter
$d_{EE}$ from the $\mu=4$ components of the amplitudes.
The parameter $d_{EE}$ enters \textit{via} one of the terms in the one-gluon
exchange vertex $R^{(1)}_{b,4}$.
Expanding this term in $\boldsymbol{p}$ and
$q$, we obtain
\begin{align}
-d_{EE}\cos(\frac{1}{2}q_4)\sum_j \gamma_j \sin q_j
\big[\sin(p_4 +q_4) - \sin(p_4)\big]
=-d_{EE} \cosh M_1 \boldsymbol{\gamma \cdot q}  q_4 + \cdots\,.
\end{align}

Then the lowest order constraint equation on $d_{EE}$ from
Eqs.~\eqref{eq:matching-vertex1} and \eqref{eq:matching-vertex2} is
\begin{align}
\bigg[\frac{\zeta e^{-M_1}\cosh M_1}{8\sinh^3 M_1}
+\frac{\zeta c_E e^{-M_1}}{8\sinh^2 M_1}
+\frac{c_{EE}}{2\tanh M_1} -d_{EE}\cosh M_1\bigg] 
\boldsymbol{\gamma \cdot q}  q_4 u(\eta,0) 
=\frac{\boldsymbol{\gamma \cdot q}  q_4 }{8m_b^3}u(\eta,0)\,, 
\label{eq:matching-dee-temporal}
\end{align}
where we gather all linearly dependent terms at 
leading order of the Taylor expansions and equate them.

For the case $\mu=i$, we obtain the lowest order 
constraint equation for $d_{EE}$ in the same manner,
\begin{align}
\label{eq:matching-dee-spatial}
\bigg[\frac{\zeta e^{-M_1}\cosh M_1}{8\sinh^3 M_1}
+\frac{\zeta c_E e^{-M_1}}{8\sinh^2 M_1}
+\frac{c_{EE}}{2\tanh M_1} -d_{EE}\cosh M_1\bigg] 
  q^2_4 \gamma_i u(\eta,0) 
=\frac{1}{8m_b^3} q_4^2 \gamma_i u(\eta,0)\,, 
\end{align}
which yields exactly the same constraint as
Eq.~\eqref{eq:matching-dee-temporal}.

We have obtained constraint equations on all the improvement parameters in the
ansatz of Eq.~\eqref{eq:field-imp}.
We have checked that all the constraint equations are 
consistent through first order in $\boldsymbol{p}$.
We expect $\mathcal{M}^{\text{lat}}_\mu$ and $\mathcal{M}_\mu$ to match through
$\mathcal{O}(\boldsymbol{p}^3)$, given the values we obtain for the 11
improvement parameters in Eq.~\eqref{eq:field-imp}.
Our results for $d_1 - d_4$ are consistent with those reported in
Ref.~\cite{Bailey:2014jga}, while for the parameters $d_B$, $d_{E}$, $d_{EE}$,
$d_{r_E}$, $d_{z_E}$, $d_{z_3}$, and $d_5$, we obtain
\begin{align}
&d_B = d_1^2 -\frac{r_s\zeta}{2(1+m_0)}
,\label{eq:result_start}\\
&d_E =\frac{1}{2m_b^2} 
-\frac{\zeta(1+m_0)(m_0^2+2m_0+2)}{[m_0(2+m_0)]^2}
+\frac{\zeta(1+m_0)(1-c_E)}{m_0(2+m_0)},
\label{eq:d_E}
\\
&d_{r_E}=-\frac{1}{8m_b^3}+\frac{r_s\zeta}{24(1+m_0)}
+\frac{\zeta c_{EE}(2+2m_0+m_0^2)}{2m_0(1+m_0)(2+m_0)}
+\frac{\zeta^2 c_{E}(2+2m_0+m_0^2)}{[2m_0(2+m_0)]^2}   
\nonumber\\
&\quad+\frac{\zeta^2(12+24m_0 +16m_0^2+4m_0^3+m_0^4)}{12m^3_0(2+m_0)^3}
-d_1 \frac{\zeta(1+m_0)[2+m_0(2+m_0)c_E]}{2m_0^2(2+m_0)^2},
\\
&d_{EE} =\frac{1+m_0}{(m_0^2+2m_0+2)}\Big[-\frac{1}{4m_b^3}
+\frac{\zeta(1+m_0)(m_0^2+2m_0+2)}{[m_0(2+m_0)]^3}
\nonumber\\
&\quad+\frac{\zeta c_E(1+m_0)}{[m_0(2+m_0)]^2}
+\frac{(2+2m_0+m_0^2)c_{EE}}{m_0(2+m_0)}\Big],
\\
&d_5 = -\frac{1}{16m_b^3}+\frac{\zeta c_B[(1+m_0)(4+6m_0+3m_0^2)\zeta -m_0^2(2+m_0)^2d_1]}
{8(1+m_0)[m_0(2+m_0)]^2} +\frac{c_3(1+m_0)}{m_0(2+m_0)}
\nonumber\\
&\quad+\frac{\zeta^2 c_E [d_1 m_0(2+m_0) -\zeta (1+m_0)]}{4[m_0(2+m_0)]^2}
+\frac{-\zeta^2 d_1 
+\zeta(1+m_0)\bigg[ 
\dfrac{(2+2m_0+m_0^2)\zeta^2}{[m_0(2+m_0)]^2} +d_B \bigg]}
{4m_0(2+m_0)},
\\
& d_{z_E} = d_{z_3}=0.
\label{eq:result_end}
\end{align}

\section{Status and outstanding issues}
\label{sec:end}

Results for the improvement parameters appearing in our ansatz for the improved
heavy quark field are contained in Eqs.~\eqref{eq:result_start}-\eqref{eq:result_end}.  
These results are obtained by matching the two- and four-quark flavor-changing current 
matrix elements at zeroth and first order in expansions in the momenta of the heavy
quarks.  
To simplify the matching of the four-quark matrix elements, we consider the 
$b\to u$ (heavy to light) transition instead of the $b\to c$
(heavy to heavy) transition, expand in the momentum transfer along the gluon
line, and match through second order in the momentum transfer.  

To cross-check the results in Eqs.~\eqref{eq:result_start}-\eqref{eq:result_end}, 
we have performed two independent sets of matching calculations.  
From the matching conditions for the two- and four-quark matrix elements, we 
find multiple independent constraints on the 11 improvement parameters. 
From these constraints we also verify that the improvement parameters in 
Eqs.~\eqref{eq:result_start}-\eqref{eq:result_end} are consistent with the
results for the improvement parameters in the OK
action~\cite{Oktay2008:PhysRevD.78.014504}.

However, several issues remain.  To summarize:  
%
%
\begin{enumerate}

\item Several of our results for the improvement parameters
  diverge as the bare mass $m_0 \to 0$.  Since we expect a smooth
  connection with the improved currents of a relativistic lattice
  theory in this limit, this divergent behavior appears suspicious, at
  least at first glance. For example, consider the parameter $d_4$
  \cite{Bailey:2014jga}.
  \begin{align}
    d_4 &= -\frac{d_1}{8M_X^2} + \frac{d_2\zeta}{4\sinh\,M_1}+ 
    \frac{3}{16}\left(\frac{1}{M_Y^3} - \frac{1}{m_b^3} \right) \,,
    \label{eq:d4}
  \end{align}
  where
  \begin{align}
    d_1 &= \frac{\zeta}{2\sinh\,M_1} - \frac{1}{2m_b}\,,
    \label{eq:d1} 
    \\
    d_2 &= d_1^2 - \frac{r_s\zeta}{2e^{M_1}}\,,
    \label{eq:d2}
    \\
    \frac{1}{M_X^2} &= \frac{\zeta^2}{\sinh^2\,M_1} 
    + \frac{2r_s\zeta}{e^{M_1}} \,,
    \\
    \frac{1}{M_Y^3} &= \frac{8}{3\sinh\,M_1}\Biggl\{2c_2 
    + \frac{1}{4}e^{-M_1}\Biggl[\zeta^2 r_s (2 \coth\,M_1 + 1) 
      \nonumber \\
      &+\ \frac{\zeta^3}{\sinh\,M_1}\Biggl(\frac{e^{-M_1}}{2\sinh\,M_1} 
      - 1 \Biggr) \Biggr] + \frac{\zeta^3}{4\sinh^2\,M_1} \Biggr\}\,.
  \end{align}
  The relationships among the rest mass, kinetic mass, and
  bare mass of the quark are
  \begin{align}
    aM_1 &= \log(1 + am_0) 
    = am_0 - \frac{1}{2} (am_0)^2 + \frac{1}{3}(am_0)^3 + \ldots
    \\
    aM_2 &= \frac{ (1 + am_0) ( (1 + am_0)^2 - 1 ) } 
    { \zeta [-r_s + r_s (1+am_0)^2 + 2 \zeta (1+ am_0) ]}
    \nonumber \\
    &= am_0 - \frac{1}{2} (am_0)^2 + (am_0)^3 + \ldots
  \end{align}
  where we set $\zeta = r_s = 1$.
  Since $m_b = M_2$, we obtain, in the limit $am_0 \to 0$,
  \begin{align}
    d_1 &= \frac{1}{4} (am_0) - \frac{3}{8} (am_0)^2 + \frac{7}{16} (am_0)^3
    + \ldots
    \\
    d_2 &= -\frac{1}{2} + \frac{1}{2}(am_0) - \frac{7}{16} (am_0)^2
    + \ldots
    \\
    d_4 &= \frac{1}{8} \frac{1}{am_0} - \frac{3}{32} - \frac{5}{32}(am_0)
    + \ldots \,.
    \label{eq:d4-te}
  \end{align}
  From the result in Eq.~\eqref{eq:d4-te}, it is clear that 
  $d_4$ has a simple pole at $am_0 = 0$.
  Hence, $d_4$ is divergent in the chiral limit.

  Similarly, we can repeat this calculation for all the improvement
  parameters $d_i$.  We find that each of $d_4$, $d_{EE}$, $d_{rE}$, and $d_5$
  has a simple pole at $am_0=0$ and so is divergent in
  the chiral limit.
  Superficially, this behavior appears to violate the first
  principle of the Fermilab formulation, because it is supposed to
  work both in the chiral limit and in the heavy quark limit
  simultaneously.
  However, we are investigating the discretization effects in the
  two-quark current matrix elements.
  We find that, as $a\to 0$ with fixed quark mass ($m_0 \ne 0$), the
  lattice artifacts vanish, and everything looks regular in this limit.
  Further progress in this direction will be reported in Ref.~\cite{
  Bailey:2017xxx}.

\item Although our results for $d_1$, $d_2$, and $d_B$ agree with
  those in Ref.~\cite{ElKhadra:1996mp}, our result for the improvement
  parameter $d_E$ differs.
  Our results for $d_E$ are given in Eq.~\eqref{eq:d_E}, and 
  those in Ref.~\cite{ElKhadra:1996mp} are
  \begin{align}
    d_E(\text{FNAL}) &= \frac{1}{2m_b^2} 
    -\frac{\zeta(1+m_0)}{m_b m_0(2+m_0)}
    +\frac{\zeta(1+m_0)(1-c_E)}{m_0(2+m_0)}
  \end{align}
  Hence, the difference is
  \begin{align}
    \Delta d_E &= d_E(\text{SWME}) - d_E(\text{FNAL})
    \nonumber \\
    &= \zeta \frac{ - (1+am_0)(2+ am_0(2+am_0)) 
      + r_s \zeta am_0 (2+am_0) + 2 \zeta^2 (1+ am_0)}
              {(am_0)^2 (2 + am_0)^2}
    \nonumber \\
    &= - \frac{1}{2 + am_0} \qquad \text{when we set $\zeta = r_s =1$.}
  \end{align}
  It converges to $\left(-\dfrac{1}{2}\right)$ in the chiral limit
  and vanishes in the heavy quark limit.
  The caveats are that $d_E(\text{FNAL})$ is obtained based on NRQCD
  power counting, for the heavy-heavy meson system, by matching the
  improved field to the canonical field with a Foldy-Wouthuysen-Tani
  transformation, while our result $d_E(\text{SWME})$ is derived using
  a momentum expansion which corresponds to HQET power counting, for
  the heavy-light meson system.
  At present, it is not clear how this difference in power counting
  has a non-trivial effect on $\Delta d_E$.
  We plan to investigate this issue in the near future.

\item Our matching condition yields a unique value for the improvement
  parameter $d_{rE}$, even though the action operator corresponding to
  this parameter is redundant.  Since the same isospectral
  transformations can be applied to the action and the currents \cite{
    Oktay2008:PhysRevD.78.014504}, a unique value for $d_{rE}$
  indicates that the operator basis given in Eq.~\eqref{eq:field-imp}
  might be incomplete.  This issue is under further investigation.

\item It has not been proved yet that the improved quark field is
  sufficient for the current improvement. The proof can be achieved
  through the HQET analysis \cite{ Kronfeld:PhysRevD.62.014505,
    Harada:2001fi,Harada:2001fj} with lattice artifacts incorporated
  in it.  We plan to do this in the near future.

\end{enumerate}

\acknowledgments
We thank Rajan Gupta and Tanmoy Bhattacharya for helpful discussion
on $V_{cb}$.
The research of W.~Lee is supported by the Creative Research
Initiatives Program (No.~20160004939) of the NRF grant funded by the
Korean government (MEST).
~J.A.B. is supported by the Basic Science Research Program of the
National Research Foundation of Korea (NRF) funded by the Ministry of
Education (No.~2015024974).
W.~Lee would like to acknowledge the support from the KISTI
supercomputing center through the strategic support program for the
supercomputing application research (No.~KSC-2014-G3-003).
Computations were carried out on the MATHEMATICA workstations at
Seoul National University.
%


\bibliography{refs}

\end{document}